\documentstyle[aps,12pt]{revtex}

\begin{document}
\draft
\author{S. N. Dolya$^{1}$ and O. B. Zaslavskii$^{2}$}
\address{{$^{1}$B. Verkin Institute for Low Temperature Physics and Engineering, 47,}%
\\
Lenin Prospekt, Kharkov 61164, Ukraine\\
E-mail: dolya@ilt.khakrov.ua\\
$^{2}$Department of Physics, Kharkov V.N. Karazin's National University,\\
Svoboda\\
Sq.4, Kharkov 61077, Ukraine\\
E-mail: aptm@kharkov.ua}
\title{General approach to potentials with two known levels}
\maketitle

\begin{abstract}
We present the general form of potentials with two given energy levels $%
E_{1} $, $E_{2}$ and find corresponding wave functions. These entities are
expressed in terms of one function $\xi (x)$ and one parameter $\Delta
E=E_{2}$-$E_{1}$. We show how the quantum numbers of both levels depend on
properties of the function $\xi (x)$. Our approach does not need resorting
to the technique of supersymmetric (SUSY) quantum mechanics but generates
the expression for the superpotential automatically.
\end{abstract}

\pacs{PACS numbers: 03.65.-w, 03.65.Ge}


\section{Introduction}

The potentials whose spectrum can be found exactly are very rare in quantum
mechanics. Meanwhile, the condition of exact solvability can be weakened:
one may demand that only for a finite part of the spectrum eigenstates and
eigenvalues be found explicitly or from a finite algebraic equation. This
opens two different possibilities. First, there exist so-called
quasi-exactly solvable (QES) systems, whose Hamiltonian can be expressed in
terms of the generators of the algebra having finite-dimensional
representation (for one-dimensional potentials the relevant algebra is $%
sl_{2}$, the corresponding generators having the meaning of the effective
spin operators) \cite{zu1} - \cite{z90}. In so doing, the dimension of the
finite subspace of the whole Hilbert space is determined by the value of the
effective spin that usually enters the QES potential as a parameter. Second,
instead of relating the dimension of the finite subspace to an underlying
structure of a Lie algebra representation, one may fix the number of known
levels ''by hands''. In the simplest case this number is equal to two, so we
deal with two-dimensional subspace. Although such a procedure makes the
underlying algebraic structure more poor, it extends considerably the set of
potentials with the known part of the spectrum.

Physical motivation for interest in potentials with two known energy levels
stems from the fact that a two-level system represents a very wide class of
models often used in solid state and nuclear physics and quantum optics. Let
us mention here only few examples: the Dicke model of interaction between
atoms and radiation \cite{dic}, Lipkin-Meshkov-Glick model of interacting
nucleons \cite{gil}, the phenomenon of macroscopic quantum tunnelling \cite
{leg}. We would like to stress that it is just the potential description of
systems with a finite number of energy levels that enabled one to give clear
and simple explanation of the phenomenon of spin tunnelling \cite{rev}.
Therefore, finding potentials, that correspond to a fixed numbers of
eigenstates, was an important step in calculation of tunneling rates in
ferro- and superparamagnets.

Meanwhile, there is also the inner motivation that stems from quantum
mechanics as such. From general viewpoint, recovering potentials from a
known set of eigenvalues is nothing else than the reduced variant of the
inverse scattering problem. As is well known, using Darboux transformation,
one can get many-soliton solutions of the Schr\"{o}dinger equation with N
energy levels, fixed in advance. Understanding, how the truncation of the
scattering data modifies the structure of the theory, could gain further
insight into the inverse scattering approach. The first necessary step here
is to find the full solution of the problem for N=2.

If N=1 (only one level is fixed), it follows from the Schr\"{o}dinger
equation that the potential is $U=E+\psi ^{\prime \prime }/\psi $, where $E$
is the value of energy, $\psi $ is a wave function. Choosing any $\psi (x)$
having no zeros at the real axis, we obtain immediately the corresponding
potential $U(x)$, regular on the real axis. We would like to stress,
however, that, in contrast to the N=1 case, when the solution of the problem
is straightforward, already for N=2 resolving this problem needed the
elaboration of different approaches discussed in literature. The existence
of exact solutions with two levels for power-like potentials was indicated
in \cite{fl2}, \cite{leach}. The rather powerful technique based on
supersymmetric (SUSY) quantum mechanics (see the review \cite{susy}) was
suggested in \cite{tk1}, \cite{tk2}. It enables one to generate the
potentials with known ground and first excited states. The aim of the
present paper is to suggest a general approach to the potentials with two
known levels valid for any n-th excited states. The corresponding method and
results turn out to be surprisingly simple and do not require sophisticated
technique (for instance, such as SUSY quantum mechanics).

\section{Basic equations}

Consider the Schr\"{o}dinger equation with the Hamiltonian $H=-\frac{d^{2}}{%
dx^{2}}+U(x)$. Let $\psi _{1}$ and $\psi _{2}$ be wave functions obeying the
Schr\"{o}dinger equation: 
\begin{eqnarray}
H\psi _{1} &=&E_{1}\psi _{1}\text{,}  \label{e1} \\
H\psi _{2} &=&E_{2}\psi _{2}\text{.}  \label{e2}
\end{eqnarray}
Then it follows from (\ref{e1}), (\ref{e2}) that 
\begin{eqnarray}
U &=&E_{1}+\frac{\psi _{1}^{\prime \prime }}{\psi _{1}}\text{,}  \label{wv1}
\\
\frac{\psi _{2}^{\prime \prime }}{\psi _{2}} &=&E_{1}-E_{2}+\frac{\psi
_{1}^{\prime \prime }}{\psi _{1}}\text{.}  \label{wv2}
\end{eqnarray}
Let, by definition, 
\begin{equation}
\psi _{2}=\xi \psi _{1}\text{.}  \label{def}
\end{equation}
Then we have for $\psi _{2}$ from (\ref{wv2}): 
\begin{equation}
\frac{\psi _{1}^{\prime }}{\psi _{1}}\equiv -\chi ^{\prime }=-\frac{(\xi
^{\prime \prime }+\Delta E\xi )}{2\xi ^{\prime }}\text{, }  \label{q}
\end{equation}
where $\Delta E=E_{2}-E_{1}$. By substitution of (\ref{def}) and (\ref{q})
to (\ref{wv2}), we obtain three equivalent forms for the potential: 
\begin{eqnarray}
U &=&E_{1}-\frac{\Delta E}{2}+\frac{3}{4}(\frac{\xi ^{\prime \prime }}{\xi
^{\prime }})^{2}-\frac{1}{2}\frac{\xi ^{\prime \prime \prime }}{\xi ^{\prime
}}+\Delta E\frac{\xi \xi ^{\prime \prime }}{\xi ^{\prime 2}}+\frac{1}{4}%
(\Delta E)^{2}(\frac{\xi }{\xi ^{^{\prime }}})^{2}\text{,}  \label{ux} \\
U &=&E_{1}-\frac{\Delta E}{2}(1-2\frac{\xi \xi ^{\prime \prime }}{\xi
^{\prime 2}})+\frac{1}{4}(\Delta E)^{2}(\frac{\xi }{\xi ^{^{\prime }}})^{2}-%
\frac{1}{2}[\xi ]_{x}\text{,}  \label{sc} \\
U &=&E_{1}+\chi ^{\prime 2}-\chi ^{\prime \prime }\text{,}  \label{su}
\end{eqnarray}
where $[\xi ]_{x}\equiv \frac{\xi ^{\prime \prime \prime }}{\xi ^{\prime }}-%
\frac{3}{2}(\frac{\xi ^{\prime \prime }}{\xi ^{\prime }})^{2}$ is Schwarzian
derivative (see, e.g., Ch. 2.7 of Ref. \cite{ba}). The wave functions of the
states under discussions are 
\begin{equation}
\psi _{1}=e^{-\chi }\text{, }\psi _{2}=e^{-\chi }\xi \text{.}  \label{wave}
\end{equation}

Eq. (\ref{ux}) gives us the general formula for the potential with two given
energy levels. It is expressed directly in terms of their values $E_{1}$, $%
E_{2}$ as parameters and one function $\xi (x)$, corresponding wave
functions are given by (\ref{q}), (\ref{wave}) and expressed in terms of the
same quantities. It is worth noting that the function $\xi (x)$ does not
enter the set of known data - rather, the freedom in its choice reflects the
fact that for two given eigenvalues there exists an infinite number of
potentials having two fixed eigenvalues. The structure of these potentials
is not arbitrary but is governed by the form of $\xi (x)$ according to (7).

Eqs. (\ref{ux}), (\ref{q}) and (\ref{wave}) constitute the main result of
this paper. It is worth stressing that the derivation of eq. (\ref{ux}) is
very simple, direct and does not need sophisticated technique, such as SUSY\
machinery. On the other hand, the potential in terms of the function $\chi
^{\prime }$ has the form (\ref{su}), typical for SUSY quantum mechanics,
automatically. In so doing, $\chi ^{\prime }$ plays the role of a
superpotential. As is well known (see, e.g., \cite{susy}), one-dimensional
quantum mechanics can always be formulated in a SUSY way. However, given a
potential, the superpotential cannot be found explicitly for a generic
model. Meanwhile, in our case we found not only the potential but the
explicit expression for the superpotential (\ref{q}) as well.

It is worth stressing that the derivation of (\ref{ux}) - (\ref{su}) relies
strongly on the successful choice of the function $\xi (x)$ that
parametrizes the family of solutions. The fact that, for given $E_{1}$, $%
E_{2}$, the ratio of two eigenfunctions determines the potential completely
generalizes the observation made in Refs. \cite{tk1}, \cite{tk2} for the
particular case when the eigenfunctions under consideration refer to the
ground and first excited states.

The formalism elaborated above for the one-dimensional Schr\"{o}dinger
equation can be also applied to the three-dimensional one for a particle
moving in a spherically-symmetrical potential $U(r)$. After the separation
of variables, the effective potential entering the radial part of the
Schr\"{o}dinger equation, is equal to $U_{ef}=U+$ $\frac{l(l+1)}{r^{2}}$.
Then, repeating calculations step by step, we obtain 
\begin{equation}
U=U^{(0)}+\lambda ^{2}\frac{\xi ^{2}}{4r^{4}\xi ^{\prime }}-\frac{\lambda
\xi }{r^{2}\xi ^{\prime }}(\frac{1}{r}+\frac{\xi ^{\prime \prime }}{\xi }%
)-\lambda \frac{\Delta E\xi ^{2}}{2r^{2}\xi ^{\prime 2}}\text{.}+\frac{%
\lambda -2l_{1}(l_{1}+1)}{2r^{2}}\text{,}  \label{ans}
\end{equation}
\begin{equation}
\lambda =(l_{2}-l_{1})(1+l_{1}+l_{2})  \label{la}
\end{equation}
and $U^{(0)}$ is expressed in terms of $E_{1}$, $E_{2}$ and $\xi $ by the
same formulas (\ref{ux}) - (\ref{su}) as in the one-dimensional case.

It is worth noting that now a new interesting possibility can arise that is
absent in the one-dimensional case: $\Delta E=0$. It becomes possible due to
the fact that two quantum states can refer to different effective potentials
($l_{1}\neq l_{2}$): we are faced with degeneracy with respect to the
angular momentum. Then the potential acquires the form 
\begin{equation}
U=E_{1}-\frac{1}{2}[\xi ]_{r}+\lambda ^{2}\frac{\xi ^{2}}{4r^{4}\xi ^{\prime
}}-\frac{\lambda \xi }{r^{2}\xi ^{\prime }}(\frac{1}{r}+\frac{\xi ^{\prime
\prime }}{\xi })\text{.}+\frac{\lambda -2l_{1}(l_{1}+1)}{2r^{2}}\text{.}
\label{deg}
\end{equation}
We will not discuss the three-dimensional case further and will concentrate
on the one-dimensional one.

\section{General properties and classification of states}

The potential $U\equiv U(E_{1},E_{2},\xi )$ possesses the symmetries that
follow directly from (\ref{ux}): 
\begin{eqnarray}
&&U(E_{1},E_{2},a\xi )=U(E_{1},E_{2},\xi )\text{,}  \label{sym} \\
&&U(E_{1},E_{2},\xi )=U(E_{2},E_{1},\xi ^{-1})\text{.}  \nonumber
\end{eqnarray}

Throughout the paper we assume that the potential $U(x)$ is regular
everywhere, except, perhaps, infinity. Then all zeros and poles of the
function $\xi (x)$ are simple - otherwise the potential $U$ would become
singular and the wave function $\psi _{2}$ would cease to be normalizable.
If the function $\xi $ has a pole at $x=x_{1}$, $\xi \approx A(x-x_{1})^{-1}$%
, one gets from (\ref{q}) that $\chi ^{\prime }\approx -(x-x_{1})^{-1}$, so $%
\psi _{1}(x_{1})=0$, $\psi _{2}(x_{1})=const\neq 0$. Therefore, every zero
of $\xi $ generates a node of the wave function $\psi _{2}$ and every pole
of $\xi $ generates a node of $\psi _{1}$.

The set of possible nodes of wave functions depends also on the behavior of
the function $\chi (x)$ in the vicinity of zeros of the function $\xi
^{\prime }(x)$ due to possible zeros of the factor $\exp (-\chi )$ in (\ref
{wave}). Let $\xi ^{\prime }(x_{0})=0$. Then, according to (\ref{q}), if $%
x\rightarrow x_{0}$, $\chi ^{\prime }\approx B/(x-x_{0})$, where $B=$ $\frac{%
(\xi ^{\prime \prime }+\Delta E\xi )_{\mid x=x_{0}}}{2\xi ^{\prime \prime
}(x_{0})}$, and the potential contains the term that behaves like $%
B(B+1)(x-x_{0})^{-2}$. The regularity of the potential entails $B=0$ or $%
B=-1.$ Consider these two cases separately.

Let, first, $B=0.$ Now the condition 
\begin{equation}
(\xi ^{\prime \prime }+\Delta E\xi )_{\mid x=x_{0}}=0  \label{reg}
\end{equation}

must hold. In so doing, the function $\chi (x)$ is regular in the vicinity
of $x_{0}$ due to the condition (\ref{reg}) and the factor $\exp (-\chi )$
cannot vanish.

Consider the case $B=-1$. Now we have 
\begin{equation}
(3\xi ^{\prime \prime }+\Delta E\xi )_{\mid x=x_{0}}=0  \label{reg2}
\end{equation}
Then $\chi ^{\prime }\approx -(x-x_{0})^{-1}$ and the functions $\psi _{1}$
and $\psi _{2}$ share the common node at $x=x_{0}$ due to the factor $\exp
(-\chi )$, as it follows from (\ref{wave}). (For example, if the potential
is even, $U(-x)=U(x)$, all wave functions of odd states vanish at $x=0$.)

As a result, we arrive at the conclusion that, if (i) $\xi (x)$ has $n_{1}$
poles and $n_{2}$ zeros, (ii) $\xi ^{^{\prime }}(x)$ has $m^{(0)}$ zeros
such that (\ref{reg}) is satisfied ($B=0$) and $m^{(-)}$ zeros such that (%
\ref{reg2}) is satisfied ($B=-1$), the function $\psi _{1}(x)$ describes the
state with the number of nodes $N_{1}=n_{1}+m^{(-)}$, while $\psi _{2}(x)$
corresponds to the state with the number of nodes $N_{2}=n_{2}+m^{(-)}$.
Therefore, the quantum number that label states is equal to $N_{1}$ for $%
\psi _{1}$ and $N_{2}$ for $\psi _{2}$ ($N_{1,2}=0$, $1$, $2...$).

It follows directly from the definition (\ref{def}): if $\psi _{1}$ has
simple zeros at $x_{i}$ and $\psi _{2}$ has simple zeros at $x_{k}$ with $%
x_{i}\neq x_{k}$, the function $\xi $ has poles at $x=x_{i}$ and zeros at $%
x=x_{k}$. However, if some $x_{i}=x_{k}$, corresponding zeros of both
functions compensate each other and this results in the fact that, if some
coefficients $B=-1$, the state labels are not determined completely by the
numbers $n_{1}$, $n_{2}$.

Let us have two fixed energy levels $E_{1}$, $E_{2}$ ($E_{2}>E_{1}$) and the
function $\xi (x)$ such that $N_{2}>N_{1}$. Then, it is the potential $%
U(E_{1},E_{2},\xi )$ for which the level $E_{1}$ belongs to the $N_{1}$-th
state and $E_{2}$ corresponds to the $N_{2}$-th state. If $N_{1}>N_{2}$, the
relevant potential is $U(E_{1},$ $E_{2},$ $\xi ^{-1})=U(E_{2},$ $E_{1,}$ $%
\xi )$, the level $E_{1}$ belongs to the $N_{2}$-th state, while $E_{2}$
corresponds to the $N_{1}$-th state. If $N_{1}=N_{2}$, the function $\xi (x)$
is not suitable for constructing $U(x)$ with two different fixed levels.
Indeed, in this case we would have had two different wave functions with the
same number of zeros corresponding to two different levels, in disagreement
with the oscillation theorem.

\section{Comparison with susy approach and tkachuk's results}

Let us introduce the function $W_{+}$ according to 
\begin{equation}
W_{+}=\frac{\delta \xi }{\xi ^{\prime }}\text{,}  \label{nw}
\end{equation}
where $\delta =\Delta E>0$.

Then, with (\ref{q}) taken into account, we obtain

\begin{eqnarray}
\chi ^{^{\prime }} &=&\frac{1}{2}(W_{+}-\frac{W_{+}^{^{\prime }}-\delta }{%
W_{+}})\equiv W\equiv \frac{W_{+}-W_{-}}{2}\text{, }U=E_{1}+W^{2}-W^{^{%
\prime }}\text{,}  \label{ss} \\
\psi _{1} &=&e^{-\int dxW}\text{, }\psi _{2}=\xi e^{-\int dxW}=W_{+}\exp [-%
\frac{1}{2}\int dx(W_{+}+W_{-})]\text{,}  \nonumber
\end{eqnarray}
where, by definition, 
\begin{equation}
W_{-}=\frac{W_{+}^{^{\prime }}-\delta }{W_{+}}\text{. }  \label{w-}
\end{equation}
Since $\psi _{1}$ must be normalizable, sign$(W_{+}(\pm \infty ))=\pm 1$.
Let $W_{+}$ have only one zero at $x=x_{0}$. If we want $W_{-}$ to be
regular at $x=x_{0}$, $W_{+}^{^{\prime }}(x_{0})=\delta $.

The formulas (\ref{ss}) (with $E_{1}=0$ and $\delta =2\varepsilon $) were
derived in \cite{tk1} by solving equations for the superpotential which
appear in the SUSY approach. In our terms, this approach deals with the
function $\xi (x)$ such, that $\xi $ has only one zero (just in the point $%
x_{0}$), $\xi ^{\prime }$ changes sign nowhere and $\xi $ does not have
poles on a real axis (otherwise they would give rise to additional zeros of $%
W_{+}$). Therefore, in the situation considered in \cite{tk1}, \cite{tk2} $%
\psi _{1}$ corresponds to the ground state and $\psi _{2}$ describes the
first excited state - in agreement with the conclusion of the previous
section of the present article. Thus, in this particular case our approach
reproduces the results of \cite{tk1}, \cite{tk2}.

\section{Illustrations. Deformations of potential leaving two levels fixed}

To illustrate the general results (\ref{ux}) - (\ref{su}), let us consider
the following example: $\xi =x^{4}+2x^{2}x_{0}^{2}-x_{1}^{4}$. The
derivative $\xi ^{\prime }=0$ at $x=0$; therefore, as is explained in the
preceding section, the corresponding example cannot belong to the set
considered in \cite{tk1}. After straightforward calculations, one obtains 
\begin{equation}
U=E_{1}+\frac{x^{2}x_{0}^{4}}{4x_{1}^{8}}+\frac{1}{4x_{1}^{4}}[A_{0}+\frac{%
A_{1}}{x^{2}+x_{0}^{2}}+\frac{A_{2}}{(x^{2}+x_{0}^{2})^{2}}]\text{,}
\label{exu}
\end{equation}
where 
\begin{equation}
A_{0}=2x_{0}^{2}(2+\frac{x_{0}^{4}}{x_{1}^{4}})\text{, }%
A_{1}=(3x_{1}^{4}+x_{0}^{4})(5-\frac{x_{0}^{4}}{x_{1}^{4}})\text{, }%
A_{2}=-(3x_{1}^{4}+x_{0}^{4})x_{0}^{2}(7+\frac{x_{0}^{2}}{x_{1}^{4}})\text{.}
\label{ai}
\end{equation}
The functions are equal to 
\begin{equation}
\psi _{1}=(x^{2}+x_{0}^{2})^{-\alpha }\exp (-\frac{x^{2}x_{0}^{2}}{2x_{1}^{4}%
})\text{, }\alpha =\frac{3x_{1}^{4}+x_{0}^{4}}{4x_{1}^{4}}\text{,}
\label{w1}
\end{equation}
\begin{equation}
\psi _{2}=\psi _{1}(x^{2}-x_{-}^{2})(x^{2}+x_{+}^{2})\text{, }x_{\pm }=\sqrt{%
x_{0}^{4}+x_{1}^{4}}\pm x_{0}^{2}\text{.}  \label{w2}
\end{equation}
It is seen from (\ref{w1}), (\ref{w2}) that $\psi _{1}$ has no nodes at the
real axis, while $\psi _{2}$ turns into zero at $x=\pm x_{-}$. Therefore, $%
\psi _{1}$ corresponds to the ground state, while $\psi _{2}$ describes the
second excited state.

As we see from (\ref{sc}), the Schwarzian derivative is an essential
ingredient of the expression for the potential under discussion. It is known
that the Schwarzian derivative is invariant with respect to the
linear-fractional transformations. Therefore, it is instructive to apply
such a transformation to the potential as a whole and look at the resulting
expression. Let us make the substitution 
\begin{equation}
\xi =\frac{c_{2}\eta +d_{2}}{c_{1}\eta +d_{1}}\text{.}  \label{c}
\end{equation}
We will use it below for generating in an explicit form rather rich families
of the potentials, corresponding to two known levels. As $[\eta ]_{x}$
remains invariant, only the part of (\ref{sc}) contains the terms
proportional to $\Delta E$ and $(\Delta E)^{2}$, changes under this
transformation. Then the potential and wave functions of the states under
discussion take the form 
\begin{eqnarray}
U &=&E_{1}-\frac{\Delta E}{2}+\frac{2\Delta Ec_{1}(d_{2}+c_{2}\eta )}{%
c_{1}d_{2}-c_{2}d_{1}}+\frac{1}{4}\frac{Y^{2}}{\eta ^{\prime 2}}-\frac{\eta
^{\prime \prime }}{\eta ^{\prime 2}}Y-\frac{1}{2}[\eta ]_{x}\text{,}
\label{uy} \\
Y &=&\Delta E\frac{(c_{1}\eta +d_{1})(c_{2}\eta +d_{2})}{%
c_{1}d_{2}-c_{2}d_{1}}\text{.}  \nonumber
\end{eqnarray}

\begin{eqnarray}
\psi _{1,2} &=&e^{-\rho }\Phi _{1,2}\text{, }\Phi _{1,2}=c_{1,2}\eta +d_{1,2}%
\text{,}  \label{wy} \\
\rho ^{^{\prime }} &=&\frac{\eta ^{\prime \prime }-Y}{2\eta ^{\prime }}\text{%
, }\chi =\rho -\ln (c_{1}\eta +d_{1})\text{.}  \nonumber
\end{eqnarray}

If $c_{2}=0=d_{1}$, one can see that $Y=\Delta E\eta $ and $%
U(E_{1},E_{2},\xi )=U(E_{1},E_{2},\eta ^{-1})=U(E_{2},E_{1},\eta $) in
accordance with (\ref{sym}).

In the limit 
\begin{equation}
c_{1}=0=d_{2}  \label{lim}
\end{equation}
we obtain the original potential $U(E_{1},E_{2},\xi )=U(E_{1},E_{2},\eta )$.

Let us assume first we have some function $\eta (x)$ characterized by the
set of numbers ($n_{1}$, $n_{2}$, $m^{(-)}$) introduced in Sec. III. The
original potential has the form (\ref{ux}) with $\xi =\eta $. Then, let us
take $\xi (x)$ according to (\ref{c}) with nonzero arbitrary coefficients $%
c_{i}$ and $d_{i}$. As the result of the transformation of (\ref{c}), each
of the aforementioned numbers can change (for example, zeros $x_{k}^{(1)}$
of the combination $c_{1}\eta +d_{1}$ generate poles of $\xi $, zeros $%
x_{i}^{(2)}$ of $c_{2}\eta +d_{2}$ correspond to zeros of $\xi $, each zero $%
x_{j}^{(0)}\neq x_{k}^{(1)}$ of $\eta ^{\prime }$ generates a zero of $\xi
^{\prime }$). Therefore, the levels $E_{1}$, $E_{2}$ which corresponded to
the $N_{1}$-th and $N_{2}$-th levels now can, in principle, correspond to
another quantum numbers ($M_{1}$, $M_{2}$). Making the transformation,
inverse to (\ref{c}), one may restore the form of the potential (\ref{ux}),
but now $\xi \neq \eta $, with $\xi $ having the form (\ref{c}), in which
coefficients under discussion play the role of parameters. Thus, we obtain a
family of deformations leaving two energy levels $E_{1}$, $E_{2}$ fixed.
These deformations can be described, on equal footing, by the deformation of
the form of the function $\xi (x)$ or of that of the potential.

For definiteness, we will choose the second possibility. If $c_{1}$, $%
c_{2}\neq 0$, one can always achieve $c_{1}=c_{2}\equiv c$ by proper
rescaling the function $\xi (x)$ that does not affect, according to (\ref
{sym}), the function $U(x)$. Then, defining 
\begin{equation}
c=2\beta \text{, }d_{1}=-\bar{\delta}-\Delta E\text{, }d_{2}=-\bar{\delta}%
+\Delta E\text{, }\gamma =\frac{(\Delta E)^{2}-\bar{\delta}^{2}}{4\beta }%
\text{,}  \label{coef}
\end{equation}
we obtain 
\begin{equation}
Y=\beta \eta ^{2}-\bar{\delta}\eta -\gamma \text{,}  \label{ny}
\end{equation}

\begin{equation}
U=E_{1}+\frac{\Delta E}{2}-\bar{\delta}+2\beta \eta -\frac{1}{2}[\eta ]_{x}+%
\frac{1}{4}\frac{(\beta \eta ^{2}-\bar{\delta}\eta -\gamma )^{2}}{\eta
^{^{\prime }2}}-\frac{\eta ^{^{\prime \prime }}}{\eta ^{^{\prime 2}}}(\beta
\eta ^{2}-\bar{\delta}\eta -\gamma )\text{,}  \label{1}
\end{equation}

Below we will see how introducing nonzero parameters $\beta $,$\gamma $
affects the potential and wave functions (\ref{wy}).

\subsection{Example 1}

Let us choose $\eta $ as a polynomial : 
\begin{equation}
\eta ^{^{\prime }}=4ax(x_{0}^{2}-x^{2})\text{, }a>0\text{, }\eta
=a(V_{0}+x_{0}^{4}-z^{2})\text{, }z=x^{2}-x_{0}^{2}\text{, }V_{0}=const.
\label{v'}
\end{equation}
Demanding that $\rho ^{^{\prime }}$ be regular at $x=0$ and at $x=\pm x_{0}$%
, we obtain from (\ref{wy}) the constraints 
\begin{equation}
\gamma =\beta a^{2}(V_{0}+x_{0}^{4})^{2}+8ax_{0}^{2}-\delta
a(V_{0}+x_{0}^{4})=\beta a^{2}V_{0}^{2}-4ax_{0}^{2}-\bar{\delta}aV_{0}\text{,%
}  \label{g}
\end{equation}
whence 
\begin{eqnarray}
V_{0} &=&-\frac{x_{0}^{4}}{2}-\frac{6}{a\beta x_{0}^{2}}+\frac{\bar{\delta}}{%
2\beta a}\text{,}  \label{x} \\
\gamma  &=&\beta ^{-1}(R-\frac{\bar{\delta}^{2}}{4})\text{, }R=\frac{(\Delta
E)^{2}}{4}=\frac{\beta ^{2}a^{2}x_{0}^{8}}{4}+2\beta ax_{0}^{2}+36x_{0}^{-4}%
\text{.}  \nonumber
\end{eqnarray}
The expression for the function $\xi $ reads 
\begin{equation}
\xi =1+\frac{2\Delta E}{A_{2}x^{4}+A_{1}x^{2}+A_{0}}\text{.}  \label{xempl}
\end{equation}
The potential can be obtained form (\ref{1}) or directly from \ref{ux}. It
has the form $U=\sum_{n=0}^{5}c_{2n}x^{2n}$, $c_{10}=\frac{(\beta a)^{2}}{64}
$.It is convenient to rescale the variable in such a way that the
coefficient in the potential $U$ at the largest power be equal to $1$. This
can be achieved by $x=\lambda y$, $\beta a\lambda ^{6}=8\omega $, where $%
\omega =1$ or $-1$. After some manipulations we get the new potential $\bar{U%
}=\lambda ^{2}U$, corresponding to levels $\bar{E}_{1,2}=\lambda ^{2}E_{1,2}$%
: 
\begin{eqnarray}
\bar{U} &=&y^{10}-6\mu y^{8}+(13\mu ^{2}+\frac{3\omega }{\mu })y^{6}-(12\mu
^{3}+22\omega )y^{4}+(4\mu ^{4}+31\mu \omega +\frac{9}{4\mu ^{2}})y^{2}
\label{u1} \\
&&+\frac{\bar{E}_{1}+\bar{E}_{2}}{2}-\frac{15}{2\mu }-6\omega \mu ^{2}\text{,%
}  \nonumber \\
\rho  &=&\frac{y^{6}}{6}-\frac{3\mu \omega y^{4}}{4}+y^{2}(\mu ^{2}\omega +%
\frac{3}{4\mu })\text{,}  \nonumber
\end{eqnarray}
\begin{equation}
(\Delta \bar{E})^{2}=16\mu ^{-2}(4\mu ^{6}+4\omega \mu ^{3}+9)\text{.}
\label{par}
\end{equation}
where $\mu =x_{0}^{2}/\lambda ^{2}$. (In fact, the formula (\ref{u1}) can be
extended to negative $\mu $ as well.) The quantities $\mu $ and $\bar{E}%
_{1,2}$ are not independent but connected by eq. (\ref{par}) that appeared
due to the condition of the regularity of the potential (\ref{reg}). It is
worth noting that the parameter $\bar{\delta}$ cancels and does not enter
the expression (\ref{u1}) due to the conditions (\ref{g}), (\ref{x}).

It follows from (\ref{wy}), (\ref{coef}) and (\ref{x}) that, up to the
constant factor, the function $\Phi _{1,2}=z^{2}-2\mu
z+q_{1,2}=(z-z_{1})(z-z_{2})$, $z\equiv y^{2}$, $z_{1,2}=\mu \pm \sqrt{\mu
^{2}-q_{1,2}}$, $q_{1,2}=\frac{\mu ^{2}}{2}+\frac{3}{4\mu }\omega \pm \frac{%
\Delta \bar{E}}{16}\omega $. Let, for definiteness, $E_{2}>E_{1}$. Consider
first the case $\omega =1$. Then after some algebra one easily finds that $%
0<q_{2}<\mu ^{2}$ and $q_{1}>\mu ^{2}$. Therefore, the function $\Phi _{1}$
does not have the nodes at the real axis and corresponds to the ground
state. The function $\Phi _{2}$ has four zeros and corresponds to the fourth
state. In a similar way, we obtain that for $\omega =-1$ the quantities $%
q_{1}<0$, $0<q_{2}<\mu ^{2}$, so the wave functions under discussion
describe the second and fourth excited states.

\subsection{Example 2}

One may exploit the ansatz (\ref{nw}) with $\xi =\eta $ for the potential (%
\ref{1}) with $\beta $, $\gamma \neq 0$. Substituting it into (\ref{wy}),
one obtains:

\begin{equation}
\rho ^{^{\prime }}=-\frac{1}{2}[\frac{(W_{+}^{^{\prime }}-\delta )}{W_{+}}-%
\frac{\bar{\delta}W_{+}}{\delta }+\frac{\beta W_{+}}{\delta }\exp (\delta
\int \frac{dx}{W_{+}})-\frac{\gamma W_{+}}{\delta }\exp (-\delta \int \frac{%
dx}{W_{+}})]\text{.}  \label{ff}
\end{equation}

Let us consider the example:

\begin{equation}
W_{+}=ax+bx^{3}\text{, }\gamma =0\text{, }\beta \neq 0\text{, }a=\delta =%
\bar{\delta}>0\text{, }b>0\text{.}  \label{ex}
\end{equation}
After simple but cumbersome manipulations we get the potential 
\begin{eqnarray}
U &=&\frac{\beta x}{x_{0}}(-\frac{x_{0}^{2}}{r}+3r+\frac{ar^{3}}{2}-\frac{%
ar^{5}}{2x_{0}^{2}})+u\text{, }r\equiv \sqrt{x_{0}^{2}+x^{2}\text{,}}\text{ }%
x_{0}^{2}\equiv \frac{a}{b}\text{,}  \label{u} \\
u &=&\frac{b^{2}}{4}x^{6}+(\frac{ab}{2}+\frac{\beta ^{2}}{4})x^{4}+\frac{%
(a^{2}-12b)}{4}x^{2}-\frac{a}{2}+\frac{3}{4(x_{0}^{2}+x^{2})}+\frac{3}{4}%
\frac{x_{0}^{2}}{(x^{2}+x_{0}^{2})^{2}}\text{.}  \nonumber
\end{eqnarray}
and wave functions 
\begin{eqnarray}
\text{ }\psi _{1} &=&\frac{(x^{2}+x_{0}^{2})^{3/4}}{(x+\sqrt{x^{2}+x_{0}^{2}}%
)^{\frac{\beta x_{0}^{3}}{16}}}e^{-\alpha }(1-\frac{\beta }{a}\eta )\text{, }%
\psi _{2}=\frac{\eta (x^{2}+x_{0}^{2})^{3/4}}{(x+\sqrt{x^{2}+x_{0}^{2}})^{%
\frac{\beta x_{0}^{3}}{16}}}e^{-\alpha }\text{, }\eta =\frac{xx_{0}}{\sqrt{%
x^{2}+x_{0}^{2}}}\text{,}  \label{exf} \\
\alpha &=&\frac{ax^{2}}{4}+\frac{bx^{4}}{8}-\frac{\beta x}{16x_{0}}%
(2x^{2}+x_{0}^{2})\sqrt{x^{2}+x_{0}^{2}}\text{.}  \nonumber
\end{eqnarray}
It is seen from (\ref{exf}) that $\eta ^{\prime }>0$ and the potential (\ref
{u}) is regular for any choice of parameters, so the conditions (\ref{reg}),
(\ref{reg2}) are irrelevant for this case. The functions $\psi _{1}$ and $%
\psi _{2}$ are normalizable, provided $\beta ^{2}<ab$. It can be readily
seen from (\ref{exf}) that the function $\psi _{2}$ has one node at $x=0$,
whereas $\psi _{1}$ turns into zero nowhere. Thus, $\psi _{1}$ and $\psi
_{2} $ correspond to the ground and the first excited states, respectively.
In the limit $\beta =0$ we reproduce the result for the example 3 of \cite
{tk1}.

\subsection{Example 3}

Let now $W_{+}=A(shx-shx_{0})$, $\gamma =0$, $\beta \neq 0$, $\delta =\bar{%
\delta}$. Then, repeating calculations for this case, we get 
\begin{eqnarray}
U &=&\frac{E_{1}+E_{2}}{2}-\frac{\delta }{2}+U_{0}(x)+\frac{\beta sh\frac{%
(x-x_{0})}{2}}{chx_{0}ch\frac{(x+x_{0})}{2}}[chx+chx_{0}-\frac{\delta
(shx-shx_{0})^{2}}{2chx_{0}}]+\frac{\beta ^{2}sh^{4}\frac{(x-x_{0})}{2}}{%
ch^{2}x_{0}}\text{,}  \label{U1} \\
U_{0}(x) &=&\frac{\delta ^{2}}{4ch^{2}x_{0}}(shx-shx_{0})^{2}-\frac{\delta }{%
2chx_{0}}(2chx-chx_{0})+\frac{1}{4}\text{,}  \nonumber \\
\psi _{1} &=&ch\frac{(x+x_{0})}{2}e^{-\alpha }(1-\frac{\beta }{\delta }\eta )%
\text{, }\psi _{2}=ch\frac{(x+x_{0})}{2}e^{-\alpha }\eta \text{, }\eta =%
\frac{sh\frac{(x-x_{0})}{2}}{ch\frac{(x+x_{0})}{2}}\text{,}  \nonumber \\
\alpha &=&(2chx_{0})^{-1}[\delta chx-\beta sh(x-x_{0})+x(\beta -\delta
shx_{0})]\text{.}  \nonumber
\end{eqnarray}
Here $U_{0}$ is the potential corresponding to the anisotropic paramagnet of
the spin $1/2$ in an oblique magnetic field \cite{zu1}. The function $\eta
^{\prime }>0$, so $\rho (x)$ is regular in any point for any choice of
parameters. The wave function is normalizable provided $-\delta
e^{-x_{0}}<\beta <\delta e^{x_{0}}$. One can easily show that it follows
from this condition that $\psi _{1}$ does not have nodes and corresponds to
the ground state, while $\psi _{2}$ has one node at $x=x_{0}$ and
corresponds to the first excited state. In the limit $\beta =0$ the example
1 of \cite{tk1} is reproduced.

\section{Concluding remarks}

Thus, in a very simple and direct approach we found a rather general
solution that gives us the structure of potentials with two known
eigenstates $E_{1}$, $E_{2}$ in terms of one function $\xi (x)$ and one
parameter coinciding with the energy difference $\Delta E=E_{2}-E_{1}$.
Moreover, we get not only the potential itself but, also (in terms of the
SUSY\ language), the superpotential. Depending on properties of the function 
$\xi (x)$ and the type of the regularity condition of the potential in the
vicinity of zeros of $\xi ^{\prime }(x)$ ((\ref{reg}) or (\ref{reg2})), one
can obtain not only the ground or first excited state but, in principle, any
pair of levels. The natural question arises whether the approach of the
present paper is extendable to the case of three (or more) levels. This
problem needs separate treatment. We hope that movement in this direction
will promote further understanding links between QES-type systems, SUSY and
the inverse scattering approaches.

One of authors (O. Z.) thanks Claus Kiefer and Freiburg university for
hospitality and acknowledges gratefully finansial support from the German
Academic Exchange Service (DAAD).

\bigskip 




%
%

%
%

\end{document}